\documentclass[aps,amssymb,amsmath,prl,reprint,superscriptaddress,noshowpacs]{revtex4-1}

\fussy

\usepackage{graphicx}
\usepackage{amsmath}
\usepackage{bm}

\usepackage{upgreek}
\usepackage[usenames]{color}
\usepackage[normalem]{ulem}

\usepackage{scalefnt}

\begin{document}
\scalefont{1.05} 

\author{Andrej Kwadrin}
\email{kwadrin@amolf.nl}
\author{A. Femius Koenderink}
\affiliation{Center for Nanophotonics, FOM Institute AMOLF, Science Park 104, 1098 XG Amsterdam, The Netherlands}

\title{Gray-Tone Lithography Implementation of Drexhage's Method for Calibrating Radiative and Nonradiative Decay Constants of Fluorophores}

\begin{abstract}
	We present a straightforward method to realize non-planar dielectric structures with a controlled height profile for use in calibration of fluorophores. Calibration of  fluorescence quantum efficiency and intrinsic radiative and nonradiative decay rates of emitters is possible by using changes in the local density of optical states, provided one can control the emitter-surface distance with nanometer accuracy. We realize a method that is accurate yet fast to implement. We fabricate PMMA wedges ($4\,$mm$\times4\,$mm$\times2\,\upmu$m) by gray-tone UV-lithography of Shipley S1813G2. Its applicability as dielectric spacer is demonstrated in Drexhage experiments for three different emitters in the visible and near-infrared wavelength regime. The decay-rate dependence of the fluorescent state of emitters on the distance to a silver mirror is observed and compared to calculations based on the local density of states. Quantitative values for (non)radiative decay rates and quantum efficiencies are extracted. Furthermore, we discuss how Drexhage experiments can help to scrutinize the validity of effective material parameters of metamaterials in the near field regime.
\end{abstract}

\maketitle

\section{Introduction}
A key contribution of photonic technology to society is the realization of light sources with desirable properties, such as controlled spectrum, brightness, improved wall-plug efficiency,  or coherence properties.  The creation of such novel light sources, i.e.,  LED's and lasers,  heavily depends on two types of innovation. These are firstly the realization of cheap  materials that have optical transitions of high efficiency and desired frequency,  and secondly the integration of these source materials with photonic structures that further improve the performance.   Among the material developments,  innovations in the last decade  range from organic dyes and light-emitting polymers in organic LEDs~\cite{Friend1999},  to the use of epitaxially grown III-V semiconductor quantum dots~\cite{Alivisatos1996} and quantum wells~\cite{Nakamura1997} in common laser diodes and LED's,  to colloidally fabricated II-VI quantum dots~\cite{Murray1993} that are size tunable throughout the entire visible spectrum.   To  improve the performance of the bare sources, or homogeneous thin layers of the luminescent materials,  many groups pursue nanophotonic techniques to improve light extraction,  as well as to  increase the spontaneous emission rate in favor of undesired nonradiative decay constants. Both, photonic crystals~\cite{Soukoulis2001} as well as patterning with arrays of plasmonic antennas~\cite{Zayats2005,Muehlschlegel2005,Muskens2007,Vecchi2009} have been shown to succesfully contribute to light extraction, and optimization of the radiative time constants~\cite{Lodahl2004}.  The key quantity that quantifies the possible improvement of the radiative decay is the local density of optical states (LDOS) enhancement that a nanophotonic structure provides~\cite{Sprik1996,Novotny}. The LDOS counts the total number of optical modes, weighted by their amplitude right at the emitter,  that are available to the source for radiative transitions, and hence directly appears in Fermi's Golden Rule for spontaneous emission~\cite{Novotny}. Optimization of LDOS is not only important for macroscopic classical light sources, such as LEDs, but is also key for quantum optical applications where a single quantum emitter is to be strongly coupled to a single optical mode~\cite{Bertet2002,Lounis2005}.

A method to reliably measure the intrinsic time constants of arbitrary emitters is of large importance both for quantifying improvements in emissive materials, and for quantifying the LDOS enhancing potential of a nanophotonic structure. For an emissive material, the challenge is to determine rapidly and accurately the intrinsic radiative rate, nonradiative rate and quantum efficiency of emitters as a tool to guide material improvement and to understand the mechanisms behind, e.g., unwanted nonradiative decay.  Conversely,  if one wishes to benchmark the local density of states improvement that a structure can intrinsically provide, it is important to probe the structure with a source that has first been quantified in terms of its radiative and nonradiative rate constants. Unfortunately, commonly used fluorescence decay measurements only provide the \emph{sum} of rates~\cite{Lakowicz2006}, while  quantum efficiency measurements are usually based on brightness comparisons. Such comparisons are prone to imprecision if one has to rely on comparison to a fluorescence quantum efficiency standard, and yield erroneous results when a sample consists of a heterogeneous ensemble in which fluorophores exhibit large brightness variations. On basis of an experiment first performed by Drexhage~\cite{Drexhage1968,Drexhage1970}, many authors have realized that intrinsic rate constants can be reliably measured by applying a known, controlled LDOS variation to an emitter~\cite{PhysRevB.55.7249,PhysRevLett.74.2459,Buchler2005}.  Drexhage studied the radiative decay rate of europium ions as a function of distance to a silver mirror. The observed variation in decay rate, stemming from interference of emitted and reflected light, can be explained by a change in the LDOS at the emitter position~\cite{Chance1978}. Since the LDOS is exactly known, the data can be quantitatively separated into an intrinsic nonradiative rate that does not vary with LDOS, and a radiative rate that does. This technique has been used for organic dyes, rare earth ions~\cite{PhysRevLett.74.2459,dood:3585},  and more  recently  for II-VI quantum dots~\cite{Leistikow2009},  III-V quantum dots~\cite{Stobbe2009,Wang2011}, and even single emitters when using a nanomechanically scanned mirror~\cite{Buchler2005}.  Unfortunately, the techniques to controllably vary distance are generally elaborate, and material specific.  For instance,  Leistikow et al.~\cite{Leistikow2009} required fabrication of a large set of samples with evaporated layers of calibrated heights.  In the case of Stobbe et al.~\cite{Stobbe2009},  a single substrate could be used, but  an elaborate  reactive ion etching step specific for III-V chemistry, and using a complicated masking procedure was required to fabricate discrete steps.    In this work we report an easily implemented  method to realize Drexhage experiments on top of arbitrary planar structures.

\begin{figure}[tbp]
\includegraphics[width=0.35\textwidth]{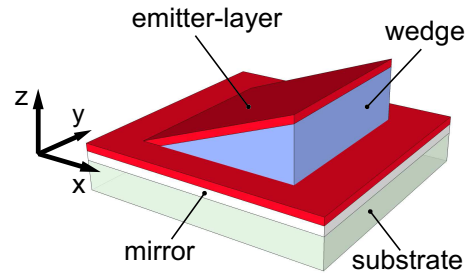}
	\caption{A wedge shaped dielectric separates a layer of emitters from a mirror. This geometry allows  to conduct a Drexhage experiment --- observing the fluorescent decay rate of emitters as a function of distance to a mirror --- on a single sample.}
	\label{fig:3Dsketch}
\end{figure}

In this paper we propose that gray-tone optical lithography~\cite{christophersen:194102} allows to attach very shallow wedges on top of arbitrary substrates.   Drexhage experiments can then be performed either by depositing the wedge on the emitter and evaporating a mirror on it, or conversely by placing the wedge on a mirror, and distributing sources on the wedge (Fig.~\ref{fig:3Dsketch}).   The key requirement for the optical wavelength regime is that the wedge has a shallow angle so that the mirror is almost parallel to the emissive substrate,  yet also  to have nanometer control over the wedge height and roughness that sets the spatial separation of the emitter and the substrate. We  fabricated wedge-shaped dielectric spacers by gray-tone lithography and performed Drexhage experiments to calibrate three different emitters: fluorescent polystyrene beads emitting at $605\,$nm, CdSeTe/ZnS (core/shell) quantum dots emitting at $800\,$nm, and Dibenzoterrylene molecules in anthracene crystals emitting at $750\,$nm. While these emitters were chosen for their promising use in nanoscale quantum optics with plasmon antennas~\cite{Akimov2007,Chang2007,Curto20082010,Koenderink2009} and metamaterials~\cite{pendry:37,Shalaev2007,Soukoulis2011}, the method is easily applied to any emitter that can be reasonably homogeneously distributed in a planar layer. The wedges can also be used with calibrated emitters,  to measure LDOS near unknown substrates, such as gratings and plasmon antenna arrays. The paper is structured as follows. In the fabrication section we discuss gray-tone UV-lithography, wedge-profile retrievement and emitter layer spin-coating steps. The experiment section covers the optical measurement procedure. Lifetime measurements and derived intrinsic rate constants for the three emitters are presented in the results section. Finally, along with our conclusion we provide an outlook.

\begin{figure*}[tbp]
\includegraphics[width=0.85\textwidth]{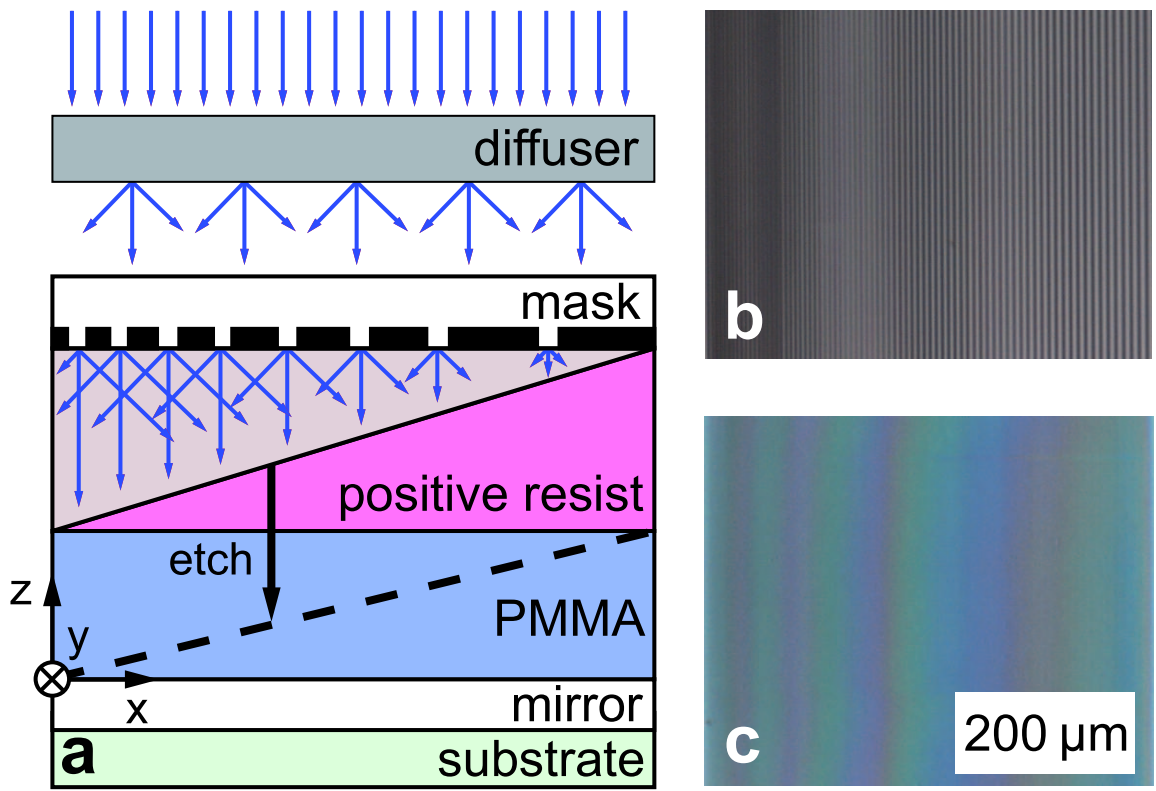}
	\caption{Diffusive UV-illumination of a 'gray-tone' chromium mask generates a wedge shaped zone with doses above the critical dose of the positive resist. After development, the wedge can be transferred into a different material, such as PMMA, via reactive oxygen ion etching (a). Differential interference contrast microscopy images of developed Shipley S1813G2 resist without diffusive element during UV-exposure, resembling the chromium mask pattern (b), and with diffusive element, showing a 'smoothed out' continuously sloping surface (c). The interference color change along x implies varying distances to the silver mirror.}
	\label{fig:Gray-tone-lithography}
\end{figure*}

\begin{figure*}[tbp]
\includegraphics[width=0.95\textwidth]{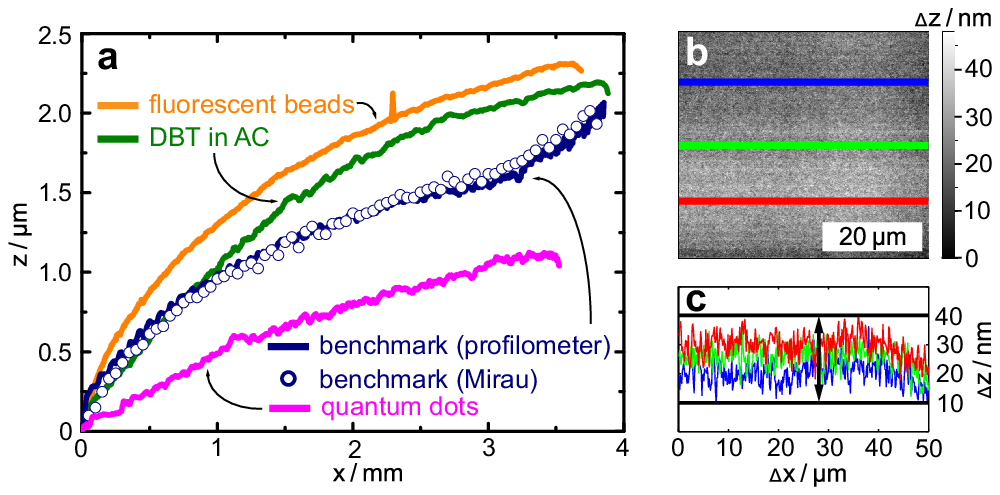}
	\caption{Panel (a):  height profiles of 4 different wedges, used for fluorescent beads, quantum dots, DBT, and a  topographic benchmark sample.  For the benchmark we also performed Mirau-interferometry (round symbols) and find good agreement  with profilometry.  The fluorescent bead and DBT in AC samples feature an additional PMMA layer underneath the S1813 UV-resist. The wedge for the quantum dot sample consists solely of S1813 (a). Panel (b): an atomic force microscopy scan  is conducted to retrieve a roughness estimate of the typical wedge surface. Line traces in  (c) show a sub-30 nm roughness as indicated by the double-arrow.}
	\label{fig:Wedge-profiles}
\end{figure*}

\section{Fabrication}
\label{sec:fabrication}

In this paper, we discuss four samples: three samples featuring emitters with specific emission wavelengths and fluorescent lifetime characteristics and one benchmark sample to check the height profile using a variety of techniques. All samples feature an optically thick ($\approx100\,$nm) layer of silver that is evaporated on silicon wafer pieces (each $\approx20\,$mm$\times20\,$mm). The silver layer, which we characterized by ellipsometry, serves as mirror for the Drexhage experiments. The essential step is the fabrication of a dielectric spacer with controlled and graded height profile using optical lithography. Since optical lithography resists tend to fluoresce themselves, we create two types of samples, namely samples with and without an intermediate, considerably less fluorescent, PMMA layer. For samples of the first type a 2\,$\upmu$m thick layer of PMMA ($M=950\,000\,$g/mol, $8\,\%$ in Anisole) is spincoated ($45\,$s at $1000\,$rpm, baking for $2\,$min at $180^{\circ}\mathrm{C}$). For all samples (with and without PMMA) we then spincoat a 2\,$\upmu$m thick layer of Shipley Microposit S1813 G2 positive UV-resist ($45\,$s at $1000\,$rpm, baking for $2\,$min at $115^{\circ}\mathrm{C}$). To define the wedge shape in the resist, we perform UV-lithography using a S\"{u}ss MJB3 mask aligner with a binary chromium mask that consists of parallel lines, at varying surface coverage similar to Christophersen et al.~\cite{christophersen:194102}. The mask for each dielectric wedge is a   $4\,$mm$\times4\,$mm  area, made up of $4\,$mm long chromium lines in the y-direction with increasing width in the x-direction from 1.5\,$\upmu$m (low end of the wedge) up to 12\,$\upmu$m (high end of the wedge). All lines are spaced in the x-direction by gaps of 3\,$\upmu$m. Thereby, the  average density continuously varies from  $33\,\%$ surface coverage to $80\,\%$ surface coverage. To generate a graded illumination, we use an opal glass diffuser (Edmund Optics, NT02-149) placed in a filter mold $\approx 1\,$cm above the sample and mask. The resist is exposed with a dose of $300\,$mJ/cm$^{2}$ and developed in Microposit MF-319 for $15\,$s (Fig.\,\ref{fig:Gray-tone-lithography}). At this point, we have a wedge profile in S1813 on top of Ag or a Ag/PMMA stack. For the two samples with PMMA, the wedge profile, defined in S1813, is transferred into the PMMA layer by reactive oxygen ion etching (Oxford Instruments Plasmalab 80+, $20\,$sccm O$_{2}$ gas flow, $50\,$W forward power, $292\,$K operating temperature). After approx.~$35\,$min of etching the S1813 is completely removed from the lower end of the wedge. The etch rate of PMMA is twice the etch rate of S1813, therefore the slope of the original profile is changed and the upper end of the PMMA wedge will be covered with residual S1813.

We characterize the wedge profiles by profilometry (KLA-Tencor Alpha-Step 500) and compare the profilometry data set of the benchmark sample to Mirau-interferometry data (Nikon 20x CF IC Epi Plan DI) (Fig.\,\ref{fig:Wedge-profiles}(a)). Since the lateral extent of the wedges is large ($4\,$mm), raw profilometry data shows curvature and roughness that are due to the scanner, rather than the sample. However, once one uses a flat optical substrate, such as an optical grade silver mirror (Thorlabs PF10-03-P01), as reference in the profilometer, a good agreement between profilometry and interferometry is obtained (Fig.\,\ref{fig:Wedge-profiles}(a), dots and curve). This supports the sole use of the profilometer as a tool to acquire (x,z)-profiles of all the other samples. Typically, the base of a wedge, in contact to the silver mirror, is $4\,$mm wide in the x- and y- (scan-) directions, as inherited from the mask. The (x,z)-profiles resemble a 'fin'-shape starting off with a steeper slope at the low end and ending with a shallower height increment per sideways displacement at the upper end of the wedges. The overall slope remains well below $1.5\,\upmu$m/mm,  meaning that the wedge angle is so shallow that effectively constant-height data can be obtained in micro-fluorescence experiments. Since neither profilometry nor Mirau-interferometry can resolve roughness reliably on lateral length scales below $1\,\upmu$m, we also performed atomic force microscopy measurements. We extract a typical surface roughness of $\Delta z/\Delta x<30\,$nm/$50\,\upmu$m (Fig.\,\ref{fig:Wedge-profiles}(b),(c)). Compared to established methods for Drexhage experiments, which are usually based on controlled stepwise RIE etching, or on controlled evaporation of different thicknesses, our method has the advantage that all desired source-mirror distances are created on a single substrate with a single simple optical lithography step. The roughness and shallow slope of the wedge imply that Drexhage experiments can be performed for wavelengths down to $\approx 500\,\mathrm{nm}$ ($<\lambda/10$ roughness).   Finally, we note that Fig.~\ref{fig:Wedge-profiles}(a) shows that while all samples have a smooth wedge,  not all wedges have the same overall height profile.  In general etching from S1813 into PMMA steepens the height profile due to a factor two difference in etch rate. In our work differences in height profile also occur due to minor variations in the spincoated thickness and in the UV exposure dose, as a result of adjustments in the mask aligner illumination alignment between runs. In this work we use individual z-profile calibrations for each wedge, though improvements in processing could render this procedure unnecessary.

\begin{figure}[tbp]
	\centering
		\includegraphics[width=0.45\textwidth]{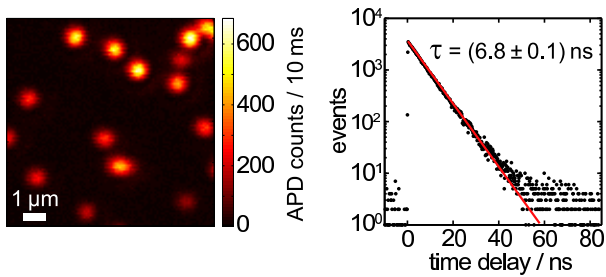}
	\caption{Typical fluorescence intensity image of beads on top of the PMMA wedge (left) and corresponding fluorescence decay histogram (accumulated events from diffraction limited bright spots in intensity image) (right). We clearly see a single exponential decay for which intensity and lifetime $\tau$ are fitted in the maximum likelihood sense assuming Poissonian counting statistics for each time bin.}
	\label{fig:beadsM039FLIMLTtrace}
\end{figure}

A succesful Drexhage experiment not only requires smooth dielectric spacers of controlled height, as characterized above, but also that the emitters can be  homogeneously dispersed on the wedge. We present measurements on three types of fluorophores. Firstly, we use fluorescent polystyrene beads (Invitrogen Fluospheres F8801), with a nominal bead diameter of $0.1\,\upmu$m containing about $10^{3}$ randomly oriented dye-molecules each, that are completely chemically shielded from the bead's environment. We estimate this number of dye molecules on the basis of intensity measurements taken in a single molecule sensitive microscope of known collection efficiency as reported by Frimmer et al.~\cite{PhysRevLett.107.123602}. Their fluorescence intensity peaks at $605\,$nm. Since S1813 wedges in themselves fluoresce in the same wavelength range, with comparable time constants,  we use a nonfluorescent PMMA wedge. To counteract agglomeration of beads we sonicate the stock solution ($2\,\%$ solids) for 2 minutes prior to mixing $1\,\upmu$l of the bead solution with $1\,$ml of deionized water. This mixture is spin-coated for $10\,$s at $500\,$rpm ($100\,$rpm/s acceleration) followed by a second spin step of $45\,$s at $1500\,$rpm ($500\,$rpm/s). From a typical confocal fluorescence microscopy image (Fig.\,\ref{fig:beadsM039FLIMLTtrace}, left) we deduce a bead concentration of $\approx 0.15\,\upmu$m$^{-2}$.

As a second emitter, we study CdSeTe/ZnS (core/shell) quantum dots (Invitrogen Qdot 800 ITK carboxyl) as they were recently proposed as light sources for plasmonic applications~\cite{Curto20082010,Akimov2007}. In comparison to more common CdSe/ZnS quantum dot nanocrystals for visible emission, these quantum dots feature longer lifetimes~\cite{vion:113120}. The quantum dots are diluted in a borate buffer solution (pH\,=\,8.0) to a molar concentration of $8\,\upmu$M. We mix $4\,\upmu$l of this solution with $1\,$ml of deionized water and spincoat a droplet of the mixture on top of the S1813 wedge-sample for $45\,$s at $4000\,$rpm ($4000\,$rpm/s). Here, the use of an S1813 wedge is justified, since the fluorescence of the quantum dots and of the polymer can be separated easily via spectral selection, and a factor $10$  contrast in fluorescence decay constants. 

As third type of emitter, we investigate Dibenzoterrylene (DBT) molecules in anthracene (AC) crystals~\cite{Jelezko1996,Toninelli:10}. DBT molecules have been found to feature great photostability and brightness at room temperature ($10^{12}$ emitted near-infrared photons before photobleaching) while also fluorescing in the near-infrared around $750\,$nm~\cite{Toninelli:10}. We dissolve $0.6\,$mg of DBT powder (Dr.\ W.\ Schmidt, PAH Reasearch Institute, Greifenberg, Germany) in $1\,$ml of Toluene. In accordance with the recipe by Toninelli~\cite{Toninelli:10},  we  use $5.3\,$mg of AC  dissolved in $2\,$ml of Diethylether and $20\,\upmu$l of Benzene. Finally, $80\,\upmu$l of DBT-solution is mixed with $0.5\,$ml of AC-solution. The resulting solution is spin-coated for $40\,$s at $3500\,$rpm ($500\,$rpm/s acceleration).

\section{Experiment}
\label{sec:experiment}

\begin{figure}[tbp]
	\centering
		\includegraphics[width=0.3\textwidth]{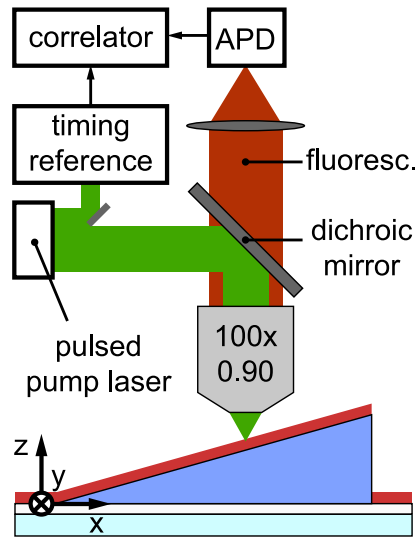}
	\caption{Sketch of the time correlated single photon counting setup. Light from a pulsed pump laser is focused onto the emitter layer. A dichroic mirror is used to separate the fluorescent light stemming from the emitters and the pump light. Depending on the emission wavelength of the emitters, appropriate long-/bandpass filters are added to the beam path. In that way only light stemming from fluorescent objects of interest is focused onto the APD. The photon arrival times with respect to the pump pulse together with the (x,y)-piezo stage position generate a fluorescent lifetime image of the sample at a certain emitter-mirror separation z.}
	\label{fig:setup}
\end{figure}

Fig.\,\ref{fig:setup} depicts the basic components of the confocal fluorescence lifetime scanning microscope, in which  we perform the Drexhage experiments. Light from the excitation laser is focused to the diffraction limit, and emitted light from fluorescent objects in the focus is collected via the same objective (Nikon 100$\times$, NA\,=\,0.90, Plan Fluor) in a confocal arrangement. The fluorescent light is separated from the excitation at a dichroic beamsplitter and passes additional long-/bandpass filters chosen according to the absorption and emission spectrum of the emitter of interest. The fluorescent beads as well as the quantum dots are pumped by a pulsed laser (Time-Bandwidth Products) operating at 532\,nm emission wavelength (green), 10\,MHz repetition rate with pulse widths $<10\,$ps. Dibenzoterrylene molecules have a 25 times higher absorption cross section in the red compared to the green part of the spectrum. Therefore, we choose a different pump source for this emitter: a pulsed laser diode (Edinburgh Instruments EPL) at 635\,nm, operated at 10\,MHz repetition rate featuring pulse widths of $<100\,$ps.

Achromatic optics focuses the fluorescent light onto a silicon avalanche photodiode (APD) (ID Quantique id100-20ULN) with an active area diameter of $20\,\upmu$m, using an effective magnification from objective to APD of $20\times$. The APD-pulses and the reference pulses from a trigger diode (green laser) or electrical trigger output (red laser) are registered by a picosecond, 16 channel, pulse correlator (Becker\&Hickl DPC-230), which records absolute  timestamps at 165\,ps resolution for each event. The sample is mounted on a (x,y)-piezo stage. By scanning the sample with respect to the objective we acquire fluorescence intensity maps of $100\times100$ pixels ($\approx10\,\upmu$m$\times10\,\upmu$m) with a pixel scan rate of 100\,Hz. For each scan we sum the single photon events of a chosen region of interest in a time-histogram. These areas are so small that no appreciable height gradient occurs in the wedge. The images hence serve to assess local homogeneity in fluorophore intensity only. For each scan we correlate the single photon detection events and laser pulses to form a fluorescence decay histogram, which we sum either over the full image or over a chosen region of interest (e.g., to select isolated beads) in the 2D scan. Different positions along the height gradient of the wedge are reached by a manually operated micromechanical stage.

\section{Results}
\label{sec:results}

\begin{figure*}[tbp]
	\centering
		\includegraphics[width=0.95\textwidth]{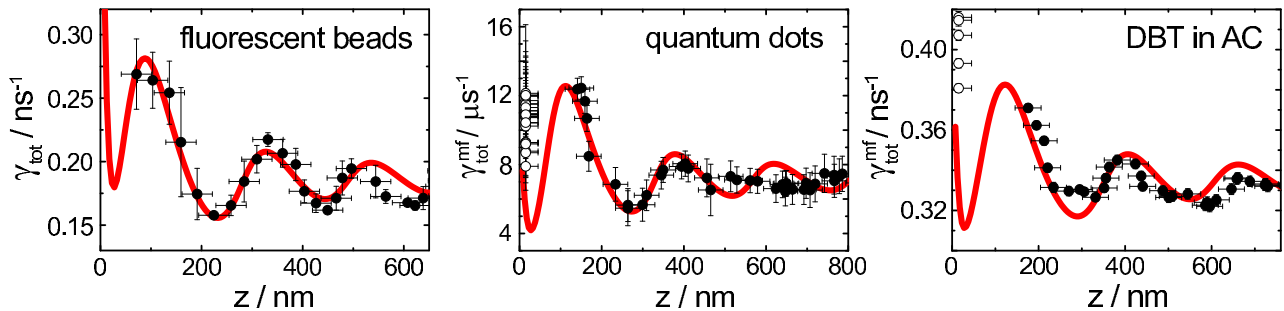}
	\caption{Fitted total decay rates $\gamma^{\mathrm{mf}}_{\mathrm{tot}}$ vs distance to the mirror z (black dots). White circles are fitted values to decay traces gathered right on top of the mirror (next to the wedge, $z\approx0$) and plotted at a fixed height offset from zero for good visibility. The red lines indicate fitted relative LDOS $\rho_{\mathrm{iso}}$ for isotropically oriented dipoles.}
	\label{fig:gamma_vs_z}
\end{figure*}

Fig.\,\ref{fig:beadsM039FLIMLTtrace}, left shows a typical scan of beads on top of the wedge aquired by fluorescence-lifetime imaging microscopy (FLIM). The intensity map shows isolated diffraction limited bright spots on a very dark background, which is consistent with having a dilute sprinkling of isolated beads. Although the slight top-bottom intensity gradient in this particular image suggests a small focus drift during the measurement, our method is robust against such drift.  Raster scanning  excludes photophysical changes in the sample, such as bleaching as artifact. Since our quantum efficiency calibration does not rely on extracting brightness, but on extracting lifetimes, we are not prone to, e.g., the small setup drifts that can cause brightness artifacts. Selecting only those events that appear in the diffraction limited bright spots we obtain a  histogram of photon counts versus arrival time after the pump pulse, i.e., a fluorescence decay curve, for all beads within the scanned image. One such histogram is shown in Fig.\,\ref{fig:beadsM039FLIMLTtrace}, right, taken from the fluorescence image Fig.\,\ref{fig:beadsM039FLIMLTtrace}, left recorded  around  position $x=340\,\upmu$m along the wedge length (i.e., $z\approx700\,$nm). For each position along the wedge profile, i.e., for each sample-source distance where we acquired a FLIM image,   we fit the observed fluoresence decay histogram with a single exponential decay. In our fit we assumed the background to be fixed (set to the mean events per  bin in the time bins with time delay $<0$, before pump pulse arrival) and fit only a decay rate and an initial amplitude. We find the most probable set of these two parameters by using the maximum-likelihood method under the assumption of Poissonian counting statistics for each time bin~\cite{Bajzer1991}. In Fig.\,\ref{fig:gamma_vs_z}, left we plot the experimentally retrieved fluorescent decay rate as a function of bead-mirror distance z. We find an increase of the decay rate up to $0.27\,$ns$^{-1}$ for $z\approx 90\,$nm. We did not resolve decay rates for values closer to the mirror which, predicted by LDOS theory, would first decrease to a minimum at around $z=25\,$nm before rising again for $z\rightarrow0$. For $z>100\,$nm, the decay rate dependancy resembles a damped oscillation around $0.18\,$ns$^{-1}$ with a periodicity of $\approx\lambda^{\mathrm{em}}_{\mathrm{bead}}/ (2n_{\mathrm{PMMA}})\approx200\,$nm as expected when considering the interference condition for emitted and reflected field amplitudes.

In order to extract a quantitative calibration of the intrinsic decay constants of the emitters, i.e., their intrinsic nonradiative decay rate and their radiative decay rate when held in vacuum, we fit the experimentally acquired total decay rate $\gamma_{\mathrm{tot}}(z)$, dependent on the emitter-mirror separation $z$, to
\begin{equation}
\gamma_{\mathrm{tot}}(z)=\gamma_{\mathrm{rad}}\cdot\rho(z)+\gamma_{\mathrm{nonrad}}
\label{eq:gamma_tot}
\end{equation}
where $\rho(z)$ equals the relative local density of states $\rho=$LDOS/LDOS$_{\mathrm{vac}}$ for a dipole above a silver mirror.
 Note that the intrinsic nonradiative decay $\gamma_{\mathrm{nonrad}}$ acts as an offset that is not affected by the mirror.  Any quenching \emph{induced by the mirror} is contained in the term proportional to the intrinsic radiative decay rate, i.e., in  $\rho$. We use established theory~\cite{Chance1978,PhysRevB.55.7249,Paulus2000,Novotny} to calculate $\rho$ on basis of the refractive indices for S1813, PMMA and the silver mirror that were acquired from ellipsometry measurements.

Since in our experiment, we acquire fluorescence intensity from an ensemble of $\approx 10^3$ dye molecules within each bead, we  assume an isotropic distribution of dipole orientations. Therefore we use the orientation averaged
\begin{equation}
\rho_{\mathrm{iso}}(z)=\frac{1}{3}\rho_{\mathrm{\perp}}(z)+\frac{2}{3}\rho_{\mathrm{\parallel}}(z)
\end{equation}
where $\rho_{\mathrm{\perp}}(z)$ and $\rho_{\mathrm{\perp}}(z)$ are the relative local density of states for dipoles oriented perpendicular and parallel to the mirror surface, respectively~\cite{Vos2009}.

\begin{figure*}[tbp]
	\centering
		\includegraphics[width=0.95\textwidth]{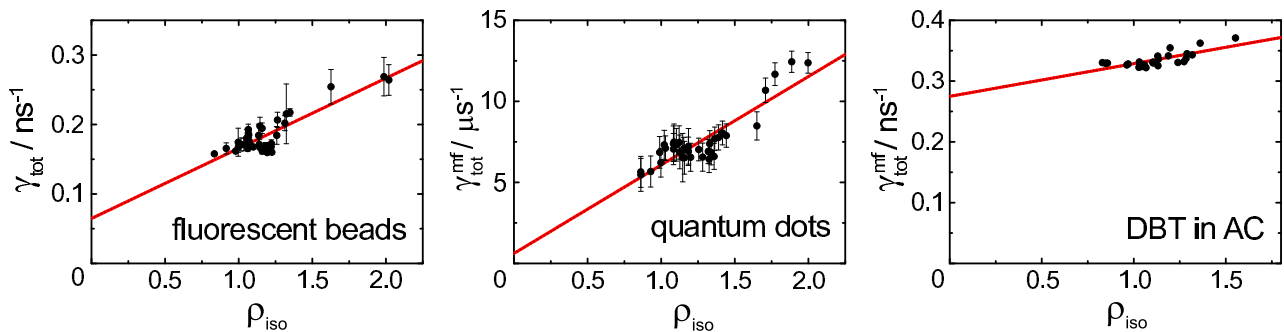}
	\caption{Total decay rates $\gamma^{\mathrm{mf}}_{\mathrm{tot}}$ vs relative LDOS $\rho_{\mathrm{iso}}$ for isotropically oriented dipoles. The red lines indicate fits according to eq.\ref{eq:gamma_tot}.}
	\label{fig:QuantumEfficiencies}
\end{figure*}

The fitted $\gamma^{\mathrm{iso}}_{\mathrm{tot}}(z)$ (calculated for dipoles embedded $10\,$nm below the PMMA/air interface to account for the bead material) is plotted as the red line in Fig.\,\ref{fig:gamma_vs_z}, left. We find fair agreement with the experimental data. For distances exceeding the vacuum emission wavelength of the emitter (due to the wedge index this encompasses several oscillations in the LDOS) the oscillation amplitude of the local density of states decreases and becomes comparable to the uncertainty in fitted total decay rates. We hence use the emitter's emission wavelength as upper bound on distances plotted. By plotting decay rate versus LDOS, instead of decay rate versus $z$, we can directly find the radiative and nonradiative decay rate from the slope and ordinate intersection of the linear dependence, respectively (Fig.\,\ref{fig:QuantumEfficiencies}). Throughout this work, error bars on rates and efficiencies result from the LDOS fit, taking into account error bars on the decay rate fitted at each vertical distance.  For the fluorescent beads, we find an intrinsic radiative decay rate of $(0.10\pm0.01)\,\mathrm{ns}^{-1}$ and a nonradiative decay rate of $(0.07\pm0.01)\,\mathrm{ns}^{-1}$.  The hypothetical  radiative decay rate  for this emitter in vacuum, extrapolates to a radiative rate of  $(0.15\,\pm0.01)\,\mathrm{ns}^{-1}$ in its PMMA host.
We extract a quantum efficiency
\begin{equation}
q.e.\equiv\gamma_{\mathrm{rad}}/(\gamma_{\mathrm{rad}}+\gamma_{\mathrm{nonrad}})
\end{equation}
of $61\%\pm7\%$ for the emitter  hypothetically in vacuum,  and around $ 70\% \pm6\%$ for the emitter embedded in its bulk host material, i.e.,  bulk PMMA.

In many applications targeting the use of metamaterials and plasmonics to control emitters, one preferentially does not use emitters with emission wavelengths at $605\,$nm, such as the beads, but rather emitters emitting further into the near-infrared. Recent reports by Curto et al.~\cite{Curto20082010} propose that CdSeTe/ZnS quantum dots emitting at around $800\,$nm are ideally suited single emitters for plasmonic applications, as they are ultrabright. The fluorescence time delay histogram retrieved from a FLIM image of an ensemble of Invitrogen Qdot 800 ITK carboxyl quantum dots does not resemble a single exponential decay. Here, we make use of the fact that an ensemble of quantum dots should be modeled with a continuous distribution of decay rates~\cite{PhysRevB.75.035329,PhysRevB.75.115302}. More precisely, the natural logarithm of decay rates $\gamma$ is assumed to be normally distributed according to the log-normal distribution
\begin{equation}
p(\gamma)=A\cdot \exp\left(\frac{\ln^2(\gamma/\gamma^{\mathrm{mf}})}{w^2}\right).
\end{equation}
The normalization constant $A$ is given by the condition $\int^{\infty}_{0}{p(\gamma)d\gamma}=1$. The dimensionless width $w$ can be rewritten as the width of the rate distribution for which $p=1/e$: 
\begin{equation}
\Delta\gamma=2\gamma^{\mathrm{mf}}\sinh{w}.
\end{equation}
Hence, the two free parameters of our fit-model are the most frequent decay rate $\gamma^{\mathrm{mf}}$, at which the log-normal distribution is centered, and the rate distribution width $\Delta\gamma$. The fitted most frequent decay rates are shown in Fig.\,\ref{fig:gamma_vs_z}, center and Fig.\,\ref{fig:QuantumEfficiencies}, center together with a fit to the LDOS for isotropically oriented dipoles. We would like to point out that as we fit the Drexhage model to just the retrieved $\gamma^{\mathrm{mf}}$, the extracted values we quote  quantify only the most frequently occurring decay rates and quantum efficiencies, and not the width of the underlying distribution. For the quantum dots we find a most frequent intrinsic radiative decay rate of $(5.46\pm0.51)\,\upmu\mathrm{s}^{-1}$, a most frequent nonradiative decay rate of $(0.62\pm0.66)\,\upmu\mathrm{s}^{-1}$ and a most frequent quantum efficiency of $90\%\pm11\%$ for the emitter hypothetically in vacuum. In its actual bulk host material, i.e., bulk S1813 we find a most frequent radiative decay rate of $(9.00\pm0.84)\,\upmu\mathrm{s}^{-1}$ and a quantum efficiency of around $ 94\%\pm7\%$.

The high quantum efficiency for these quantum dots makes them exceptionally promising for plasmon quantum optics, since operating at $800\,\mathrm{nm}$ optimizes the emission frequency to be at the intersection of low plasmon loss, yet efficient silicon detection. The high quantum efficiency is surprising given that the solvent/ligand exchange to aqueous condition, and operation of quantum dots unprotected against oxygen as in our experiment usually adversely affects the photophysical properties.  Here we note that measuring quantum efficiency via modulation of radiative lifetime selectively measures the quantum efficiency of the ensemble of dots that  radiate, i.e., that are not temporarily or permanently dark due to (photo)chemical processes such as bleaching or blinking.  This property should be contrasted to   absorption/emission brightness measurements that report  quantum efficiency as the fraction of absorbed photons that is converted into radiated photons by an ensemble of nanocrystals. In such measurements a mixture of a dark subensemble that absorbs but does not emit with a subensemble of  unit-efficiency  emitters is indistinguishable from a homogeneous sample of emitters with below-unit quantum efficiency.  The conclusion from our measurement is that those dots that  are not (photo)chemically altered by an oxygen-rich or aqueous environment, retain high quantum efficiency. A similar conclusion that the quantum efficiency of those dots that radiate is much higher than the ensemble efficiency obtained from a absorption/emission brightness measurements was also reached by Leistikow~\cite{Leistikow2009}, for CdSe quantum dots emitting around $600\,\mathrm{nm}$, with efficiencies between 66\% and 89\%.  Evidently, the CdSeTe/ZnS core shell dots emitting at $800\,\mathrm{nm}$ have a 6-fold longer lifetime compared to CdSe quantum dots, consistent with observations by Vion et al.~\cite{vion:113120}.  It is remarkable that the (bright) $800\,\mathrm{nm}$ dots in our work manage to retain a very high quantum efficiency, despite this 6-fold longer lifetime~\cite{Leistikow2009}.

An interesting alternative to the quantum dots, which while efficient, have a slow and non-single exponential decay, could be DBT with an emission wavelength of $\approx 750\,$nm~\cite{Jelezko1996,Toninelli:10}. DBT was recently reported to be ultrastable and ultrabright as an emitter at room temperature~\cite{Toninelli:10}. In contrast to Toninelli's observations  on single molecules, we observe a non-single exponential decay trace for ensembles of DBT molecules. Therefore, we apply the same analysis scheme as for the quantum dots: log-normally distributed decay rates~\cite{PhysRevB.75.035329}. In this manner we find a most frequent intrinsic radiative decay rate of $(0.05\pm0.01)\,\mathrm{ns}^{-1}$, a most frequent nonradiative decay rate of $(0.27\pm0.01)\,\mathrm{ns}^{-1}$ and a most frequent quantum efficiency of $16\%\pm3\%$,  when calculating with the radiative rate extrapolated to vacuum. Decay rate data and LDOS fit for DBT are shown in Fig.\,\ref{fig:gamma_vs_z}, right and Fig.\,\ref{fig:QuantumEfficiencies}, right. Since DBT is always used in anthracene, it is useful to extract the quantum efficiency in bulk anthracene. Correcting for the index of refraction of anthracene, we find a most frequent radiative decay rate of $(0.09\pm0.02)\,\mathrm{ns}^{-1}$ and a quantum efficiency around $24\%\pm 4\%$.

The moderate to low quantum efficiency that our measurement retrieves for DBT is surprising, since Toninelli et al.~\cite{Toninelli:10} have reported very high count rates from single DBT molecules in anthracene at room temperature. In saturation, they reported  detecting photons at a rate of  up to 0.5\% of the inverse lifetime.  A quantum efficiency of around 25\%  is within, but at the low end of,  the range of values that are consistent with  the observation of Toninelli et al.~\cite{Toninelli:10}, depending on the actual microscope collection efficiency (state of the art: a few percent at $800\,\mathrm{nm}$). We note that the observations of Toninelli et al.~\cite{Toninelli:10} are strictly for individual molecules selected to be ideal in the sense of being long lived, which likely selects molecules from those 10\% best incorporated in the anthracene matrix. While agreement with LDOS theory for isotropic dipole orientations is good, we found poor agreement of decay rate data to LDOS lineshapes for specific dipole orientations, despite reports of preferential orientation for this system~\cite{Toninelli:10}. This further highlights that we probe a heterogeneously distributed ensemble of emitters, as opposed to selecting particular emitters as in single molecule experiments~\cite{Toninelli:10}. From observation of the film quality, we note that it was difficult to obtain homogeneous anthracene crystal films throughout the entire wedge length, both on the wedge material as well as on clean cover slips.

Tab.~\ref{tab:TableOfEmitterProperties} summarizes the retrieved radiative and nonradiative decay constants and respective quantum efficiencies for all three fluorophores.

\begin{table*}[tbph]
	\centering
		\begin{tabular}{|c|c|c|c|}
			\hline
				 & Invitrogen Fluospheres F8801 & Invitrogen Qdot 800 ITK carboxyl & DBT in AC \\ \hline
				$\gamma^{\mathrm{mf}}_{\mathrm{rad,host}}$ & $(0.15\pm0.01)\,\mathrm{ns}^{-1}$ & $(9.00\pm0.84)\,\upmu\mathrm{s}^{-1}$  &  $(0.09\pm0.02)\,\mathrm{ns}^{-1}$\\ \hline
				$\gamma^{\mathrm{mf}}_{\mathrm{nonrad}}$ & $ (0.07\pm0.01)\,\mathrm{ns}^{-1}$& $(0.62\pm0.66)\,\upmu\mathrm{s}^{-1}$  &  $(0.27\pm0.01)\,\mathrm{ns}^{-1}$\\ \hline
				$q.e.^{\mathrm{mf}}_{\mathrm{host}}$ & $71\%\pm6\%$ & $94\%\pm7\%$ & $24\%\pm4\%$ \\ 
			\hline
		\end{tabular}
	\caption{Table of emitter properties retrieved from our Drexhage experiments. As we assumed a log-normal distribution of decay rates for the probed Invitrogen Qdot and DBT ensembles, the stated decay rates and quantum efficiencies are the ones which are most-frequent (superscript 'mf'). Radiative decay rates and quantum efficiencies are quoted for the emitters embedded in their respective host medium: PMMA (Fluospheres), S1813 (quantum dots), and anthracene (DBT).}
	\label{tab:TableOfEmitterProperties}
\end{table*}

\section{Conclusions and outlook}
\label{sec:outlook}

In conclusion, we have demonstrated that gray-tone UV-lithography~\cite{christophersen:194102} provides a facile method to create samples to calibrate ensembles of emitters. The essential steps of this method are (1) gray-tone lithography to create S1813 or PMMA polymer wedges on  flat reflective substrates, such as an Ag or Au mirror,  (2) a homogeneous dispersal method to distribute fluorophores on the wedge, and (3)  lifetime measurements along the length of the wedge to effectuate Drexhage experiments~\cite{PhysRevLett.74.2459}.  Previous methods rather used controlled stepped etching into substrates that had the light sources embedded in them,  i.e.,  a material-specific technique suited for III-V sources~\cite{Stobbe2009},  or depended on the creation of a large but discrete set of substrates with different deposition thicknesses of dielectric spacer layer~\cite{Leistikow2009}.  Instead, our method allows to create a continuous wedge on a single substrate. Our method can thus be of large use for calibrating a range of fluorophores for applications ranging from quantum optics, to organic and inorganic light emitting diodes.   We note that a large range of variations is easily implemented.  For instance, if surface chemistry means it is advantageous to first deposit the fluorophores on glass, then create a wedge, and then deposit a mirror,  this is equally easily implemented.
Beyond calibration of unknown emitters using a known LDOS, the same method is also of large interest to do the reverse, i.e., to measure an \textit{unknown} LDOS using a calibrated emitter.  Indeed,  in plasmonics, metamaterials, and the new field of metasurfaces, one frequently encounters questions that revolve around the LDOS at controlled distance.  Consider for instance the optimization of light emitting diodes by plasmon particle array surfaces~\cite{Zayats2005,Muskens2007,Vecchi2009} or extraordinary transmission gratings~\cite{Ebbesen1998}.  The question at which distance one should  optimally place the emitters to both enhance outcoupling,  radiative rate, and quantum efficiency is of key importance, yet difficult to address experimentally.   Likewise,  the fundamental  study of effective medium parameters of metamaterials come to mind. Photonic metamaterials are artificial materials with periodic arrangements of subwavelength scatterers aimed at arbitrary control of permittivity $\epsilon$ and permeability $\mu$  to realize perfect lenses and invisibility cloaks via transformation optics~\cite{PhysRevLett.85.3966,Valentine2009,Ergin2010,Zentgraf2010}.  For a wide range of metamaterials $\epsilon$ and $\mu$ have been reported~\cite{Shelby2001,Dolling2006,Valentine2008}, showing that it is indeed possible to obtain, e.g., effectively negative $\mu$.   Effective medium constants are commonly retrieved from \emph{far field} experiments.  Since metamaterials have intrinsic near field benefits, it is interesting to test whether these effective material parameters retrieved from far field measures such as transmittance and reflectance, are still valid in the near field.   Using our wedge technique one can continuously sweep source height through the near field zone to examine  the transition from resolving individual building blocks to resolving just the effective parameters~\cite{PhysRevA.71.011804,PhysRevA.68.043816}.   The polymer wedges that we fabricate can indeed also be readily made on patterned surfaces, thereby opening the road to controllably  vary the near-field spacing  and measure the metamaterial LDOS.  This method is complementary to elaborate near-field scanning methods~\cite{PhysRevLett.107.123602}.  The  loss of lateral resolution  in our method compared to near-field scanning is offset by the ease of use of the gray-tone lithography wedge technique, and the fact that in many applications (e.g., plasmon enhanced LEDs)  the only useful quantity would anyway be of ensemble averaged nature.

\begin{acknowledgements}
We thank Martin Frimmer for initial help with lifetime measurements, Gijs Vollenbroek for assistance with sample fabrication, and Dr.~Toninelli for kindly sharing her experience with depositing DBT in anthracene. This work is part of the research programme of the Foundation for Fundamental Research on Matter (FOM), which is part of the Netherlands Organisation for Scientific Research (NWO). AFK acknowledges a NWO-Vidi fellowship.
\end{acknowledgements}

\bibliography{AKwadrin_AFKoenderink_Calibrating_fluorophores}

\end{document}